\documentclass[aps,prb,showpacs,reprint,superscriptaddress,dvipdfmx]{revtex4-1}

\usepackage[dvipdfmx]{graphicx}
\usepackage{color}
\usepackage{latexsym}

\def\cp3s{CePt$_3$Si}
\def\upa{UPd$_2$Al$_3$}
\def\tn{$T_{\rm N}$}
\def\tsc{$T_{\rm sc}$}
\def\BB{{\textit{\textbf{B}}}}

\def\Ba{{\textit{\textbf{B}}}${\parallel}$[1~0~0]}
\def\Bc{{\textit{\textbf{B}}}${\parallel}$[0~0~1]}
\def\bc2{{\textit{\textbf{B}}}$_{\rm c2}$}
\def\mub{${\mu}_{\rm B}$}

\def\qaf{{\textit{\textbf{q}}}$_{\rm AF}$}

\def\degs{$^{\circ}$}

\begin{document}

\preprint{APS/123-QED}

\title{Magnetic-field enhanced aniferromagnetism in non-centrosymmetric heavy-fermion superconductor CePt$_3$Si}

\author{K. Kaneko}
\email{kaneko.koji@gmail.com}
\affiliation{Quantum Beam Science Center, Japan Atomic Energy Agency, Tokai, Naka, Ibaraki 319-1195, Japan}
\affiliation{Max-Planck-Institut f{\"u}r Chemische Physik fester Stoffe, 01187 Dresden, Germany}

\author{O. Stockert}
\affiliation{Max-Planck-Institut f{\"u}r Chemische Physik fester Stoffe, 01187 Dresden, Germany}

\author{B. F{\aa}k}
\affiliation{SPSMS, UMR-E 9001, CEA-INAC/UJF-Grenoble 1, 38054 Grenoble, France}

\author{S. Raymond}
\affiliation{SPSMS, UMR-E 9001, CEA-INAC/UJF-Grenoble 1, 38054 Grenoble, France}

\author{M. Skoulatos}
\affiliation{Helmholtz-Zentrum Berlin f\"ur Materialien und Energie, 14109 Berlin, Germany}
\affiliation{Laboratory for Neutron Scattering, Paul-Scherrer Institut, 5232 Villigen, Switzerland}

\author{T. Takeuchi}
\affiliation{Low Temperature Center, Osaka University, Toyonaka 560-0043, Japan}

\author{Y. \=Onuki}
\affiliation{Faculty of Science, University of the Ryukyus, Nishihara, Okinawa 903-0213, Japan}

\date{\today}

\begin{abstract}
The effect of magnetic field on the static and dynamic spin correlations in the non-centrosymmetric heavy-fermion superconductor CePt$_3$Si was investigated by neutron scattering.
The application of a magnetic field {\BB} increases the antiferromagnetic (AFM) peak intensity. 
This increase depends strongly on the field direction: for \Bc\ the intensity increases by a factor of 4.6 at a field of 6.6~T, which corresponds to more than a doubling of the AFM moment, 
while the moment  increases by only 10\% for \Ba\ at 5~T.
This is in strong contrast to the inelastic response near the antiferromagnetic ordering vector, where no marked field variations are observed for {\Bc} up to 3.8~T.
The results reveal that the AFM state in {\cp3s}, which coexists with superconductivity, is distinctly different from other unconventional superconductors.

\end{abstract}

\maketitle

The characterization of the static and dynamic magnetic response is of particular importance for the  study of the interplay between magnetism and superconductivity in unconventional superconductors.
Unconventional superconductivity is mostly realized  in the vicinity of a magnetic quantum critical point, which suggests that spin fluctuations play a vital role in forming the Cooper pairs.
The microscopic coexistence of long-range magnetic order with a superconducting (SC) state is rarely seen,  
except in a few U-based heavy-fermion superconductors, such as UPd$_2$Al$_3$.\cite{Geibel1991} 

{\cp3s} is particularly interesting in this context, since, in addition to its non-centrosymmetric crystal structure, 
it is a unique example among Ce-based heavy-fermion compounds where superconductivity is realized in the antiferromagnetic (AFM) state in a stoichiometric compound already at ambient pressure.\cite{Bauer2003} 
{\cp3s} crystallizes in the tetragonal CePt$_3$B-type structure (space group No. 99, $P4mm$) with $a$ = 4.072~{\AA} and $c$ = 5.442~{\AA}.  
High-quality single crystals of CePt$_3$Si exhibit a SC transition at {\tsc}=0.45\,K,  below the AFM ordering temperature at {\tn}=2.2~K.\cite{Takeuchi2007} 
The AFM order of {\cp3s} is described by a simple commensurate propagation vector {\qaf}=(0\,0\,$\frac{1}{2}$),
and a reduced ordered moment of 0.17~{\mub} lying in the basal plane.\cite{CePt3Si_01}
The microscopic coexistence of the long-range AFM order with the SC state is shown consistently by neutron scattering,\cite{CePt3Si_01} NMR,\cite{Yogi2004} and ${\mu}$SR\cite{CePt3Si_AAmato_1} techniques.
Despite of their coexistence, magnetism and superconductivity are found to be weakly coupled in {\cp3s}.
The SC phase in {\cp3s} persists beyond the critical pressure where AFM order vanishes.\cite{Tateiwa2007,Takeuchi2007,Nicklas2010}
In addition, both static and dynamic spin correlations around {\qaf} are almost unchanged on passing through {\tsc}.\cite{Fak2008} 
This is in clear contrast to {\upa}, where significant changes in the magnetic spectral weight are observed at {\tsc} around {\qaf}.\cite{Bernhoeft1998,Metoki1998}

The application of a magnetic field is a powerful approach to unveil hidden magnetic properties and their relationship with other order parameters.
Indeed, magnetic field may induce novel ordered states such as the $Q$-phase in superconducting CeCoIn$_5$\cite{Bianchi2003,Kenzelmann2008} 
and the antiferroquadrupolar order in heavy-fermion superconductor PrOs$_4$Sb$_{12}$\cite{PrOs4Sb12_07,Kohgi2003}.
Indeed, additional field-induced phase boundaries are revealed inside the AFM ordered state in {\cp3s}.
The magnetostriction measurement shows temperature independent anomalies around 4~T for field applied both parallel and perpendicular to the basal plane.\cite{Takeuchi2007,Takeuchi2004} 
This additional anomaly inside the AFM state is confirmed by the specific heat for {\Bc}  as a peak splitting between 3 and 4~T.\cite{Takeuchi2007} 
However, the origin of these anomalies remains unclear.

In this paper, we use neutron scattering in magnetic fields to reveal the unique characteristics of the antiferromagnetism in {\cp3s}.
We show that the AFM moment is enhanced when a magnetic field is applied parallel to both [0~0~1] and [1~0~0]. The effect is particularly strong for {\Bc}, where the ordered moments more than double at 6.6~T.
This drastic field-induced  modification of the elastic scattering is not accompanied by any visible variation in the inelastic response around {\qaf}.
This observation suggests unusual hidden nature of the AFM state in superconducting {\cp3s}.

\begin{figure}[t]
	\begin{center}
		\includegraphics[width=8.5cm]{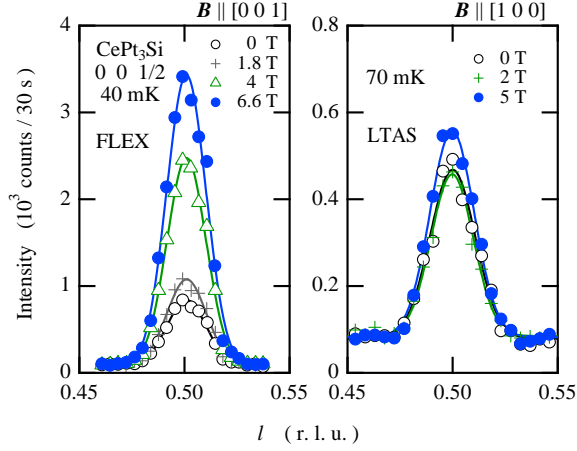}
	\end{center}
	\vspace{-5mm}
	\caption{(Color online) Scans along the $l$ direction of the 0~0~$\frac{1}{2}$ reflection taken at several magnetic fields applied for both {\Bc} and [1~0~0].}
	\label{f1}
	\vspace{-5mm}
\end{figure}%

Neutron scattering experiments on single crystalline {\cp3s} were carried out on three cold neutron triple-axis spectrometers, LTAS, FLEX and IN14.
The high quality of the single crystals used in this work were confirmed by sharp jumps in the heat capacity at $T_N=2.2$~K and at $T_{sc}=0.45$~K. 
The largest sample had a cylindrical shape with 6~mm diameter and 10~mm height, and was mounted in a dilution refrigerator with (0~$k~l$) as the horizontal scattering plane.
Elastic neutron scattering experiments in magnetic fields up to 5~T for both {\Ba} and [0~0~1] using vertical and horizontal field magnets, respectively, were carried out on LTAS at JRR-3 in Tokai.
Data at higher magnetic fields up to 6.6\,T for {\Bc} were taken on FLEX with the horizontal field magnet HM-3 at the BER-II reactor in Berlin.
Both these experiments used a final neutron wave vector of $k_f$=1.55~{\AA}$^{-1}$, which gives an energy resolution better than 0.2~meV FWHM (full width at half maximum).
Inelastic neutron scattering spectra in fields up to 3.8~T for {\Bc} were recorded on IN14 at the Institut Laue-Langevin, Grenoble, with a horizontal field magnet having a wide-angle window.
To achieve higher energy resolution, a set up with $k_f$=1.05~{\AA} without collimators was employed resulting in an energy resolution better than 60~${\mu}$eV FWHM. 
A cooled Be-filter was placed in the neutron beam before (LTAS, FLEX) or after (IN14) the sample to eliminate higher order contaminations. 

\begin{figure}[!!tttt]
	\begin{center}
		\includegraphics[width=9cm]{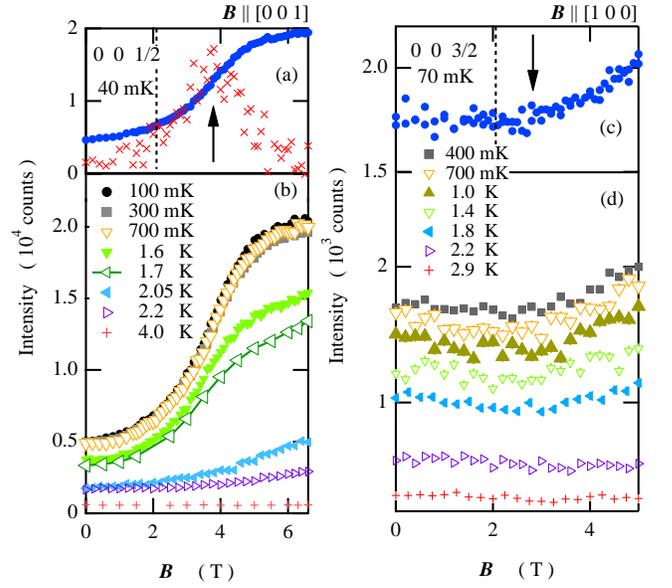}
	\end{center}
	\vspace{-5mm}
	\caption{(Color online) AFM peak intensities as a function of applied field for {\Bc} (a--b) and {\Ba} (c--d) measured at several temperatures.
	Data taken at base temperature are shown in the upper panel (a,c), in which dotted lines and arrows indicate {\bc2} and inflection points, respectively. 
	Panel (a) shows the derivative as well (\textcolor{red}{${\times}$}). Statistical errors of the intensities are smaller than the symbol size}
	\label{f2}
\end{figure}%

First, the effect of magnetic field on the AFM Bragg peak is described.
Figure~\ref{f1} shows scans along $l$ through the  0\,0\,$\frac{1}{2}$ reflection measured under several magnetic fields for {\Bc} and [1~0~0] below 100~mK.
The AFM intensity increases  with increasing field for both applied field directions. 
The increase is small for \Ba\ while it is very strong for \Bc. 
No shift or broadening of neither the magnetic (see Fig.~\ref{f1}) nor the nuclear Bragg peaks (not shown) were observed.
This observation allowed us to study the magnetic field response by simply measuring the peak intensity at the nominal peak position. 
The field dependence of the AFM peak intensities for both applied field directions below 4\,K is summarized in Fig.~\ref{f2}.
The result for {\Bc} at base temperature [Fig.\ \ref{f2}(a)] demonstrates the continuous increase of the AFM intensity with field from zero up to 6.6~T. 
The total gain in magnetic intensity reaches a factor of 4.6, after subtraction of the field-independent background determined at $T=4$~K.
In addition, the magnetic intensity displays an inflection point  in the field dependence,  an observation corroborated by taking the field derivative of the intensity as shown in the same figure.
The derivative at $T=40$~mK shows a clear maximum near 4~T.
In contrast, no anomaly was found around the superconducting transition at $B_{\rm c2}\approx2$~T.
At higher temperatures, almost identical field responses were obtained up to 700~mK as displayed in Fig.\ \ref{f2}(b).
Whereas the absolute intensity decreases with increasing temperature, the field-induced increase persists up to {\tn}.
The inflection point in the intensity is observed up to temperatures of 1.7~K but not beyond. 
No hysteresis was observed for any of the isothermal intensity versus field curves.

In contrast, the magnetic response for an in-plane magnetic field {\Ba} is different from that for {\Bc}, 
although the AFM intensity increases for both field directions.
The magnetic field dependence of the 0~0~3/2 peak intensity at $T=70$~mK is plotted in Fig.\ \ref{f2}(c).
A distinct difference from {\Bc} can already be seen in the initial slope. 
The intensity first decreases slightly from zero field up to 3~T.
Above 3~T, the field response changes its slope to positive, and keeps increasing up to a field of 5\,T.
The gain from 0 to 5~T for {\Ba} is only 20\,\%, i.e. much smaller than what is observed for {\Bc}.
The AFM intensity isotherms above 400~mK  are summarized in Fig.\ \ref{f2}(d).
The overall field response including the inflection around 3~T is almost temperature independent up to 1.8~K without any field hysteresis,  and therefore qualitatively similar to {\Bc}.

\begin{figure}[tttt]
	\begin{center}
		\includegraphics[width=8cm]{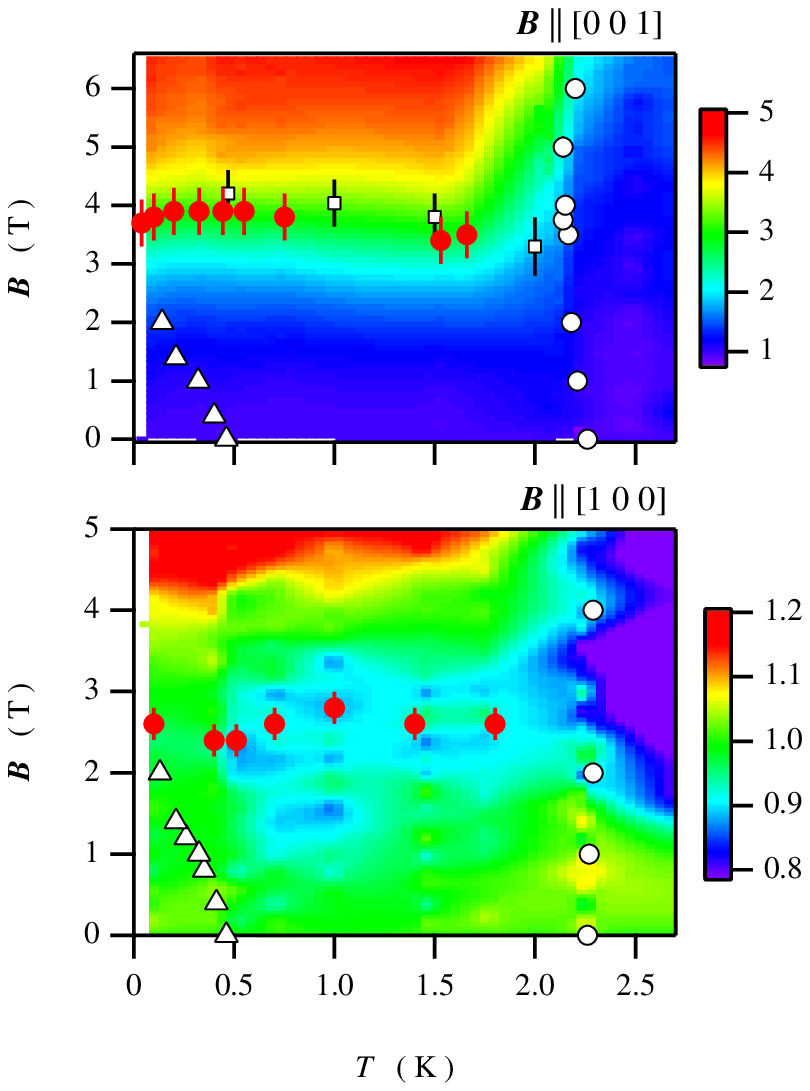}
	\end{center}
	\vspace{-5mm}
	\caption{(Color online) Magnetic $B$-$T$ phase diagram of {\cp3s} for (a) {\Bc} and (b) {\Ba}.
	Normalized gain of the AFM intensity represented by Eq.~\ref{eq1} is superposed upon the diagrams as a colour map.
	Closed (red) circles represent inflection fields in the AFM intensity isotherms, and open symbols corresponds to 
	boundaries of the superconducting phase (${\triangle}$), the AFM order (${\circ}$), and an additional unidentified phase (${\Box}$) taken from ref. \onlinecite{Takeuchi2007}.}
	\label{f3}
\end{figure}%

To summarize these findings, the temperature-field response of the AFM peak intensity is plotted as a color map in the $B$-$T$ phase diagrams in Fig.~\ref{f3} for both field directions.
The quantity shown is the normalized intensity gain at each temperature and field defined as
\begin{equation}
	g(B, T)=(I(B, T)-I_{\rm bg})/(I(B=0, T)-I_{\rm bg})
	\label{eq1}
\end{equation}%
where $I(B=0,T)$ is the intensity in zero field for each temperature and $I_{\rm bg}$ is the field-independent background taken from data well above {\tn}, at $T=4$~K and 2.9~K for {\Bc} and [1~0~0], respectively.
The field dependence is essentially temperature independent below $T=1.5$~K, 
which implies that the intensity gain has no relationship to the {\bc2} superconducting field (shown by ${\triangle}$ symbols in Fig.\ \ref{f3}), which is isotropic and temperature dependent. 
The ratio of the intensity gain for the two field directions is 3.8, indicating a large field anisotropy.  This is in clear contrast to the phase diagram, which is almost independent of the field direction. 
The inflection points in the isotherms around 4 and 3~T for {\Bc} and [1~0~0], respectively, are temperature independent and coincide with the reported phase boundaries inside the AFM phase.\cite{Takeuchi2007}

\begin{figure}[tttt]
	\begin{center}
		\includegraphics[width=7cm]{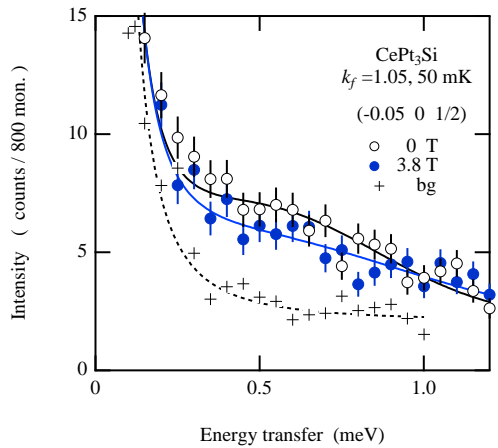}
	\end{center}
	\vspace{-5mm}
	\caption{(Color online) Inelastic neutron scattering spectra at \textbf{\textit{Q}}=(-0.05\,0\,$\frac{1}{2})$ (near {\qaf}) taken for magnetic fields  \Bc\ of 0 and 3.8~T at $T=50$~mK. Data obtained at (0~0~0.75) is taken as background.
	Lines are a guide to the eyes.}
	\label{f5}
\end{figure}%

We now discuss the possible origin of the observed enhancement in the AFM intensity.
The observed magnetic intensity is dominated by two key factors, the direction and the size of the ordered moment.
Since neutron scattering only observes magnetic moment components perpendicular to the momentum transfer, 
the intensity observed for 0 0 $l$ reflections is related to the basal-plane component of the staggered AFM moment. 
On the assumption that the zero-field moment is in the basal plane,\cite{CePt3Si_01} 
which is supported by symmetry analysis,\cite{Fak2008}
any moment reorientation would lead to a loss of the AFM intensity for these reflections. 
If an observed intensity increase arises only from moment reorientation,
the AFM moment must be canted out of the basal plane more than 60{\degs} in zero field, 
which cannot explain the observed AFM intensity in ref.\cite{CePt3Si_01}.
Hence, an increase of the intensity is necessarily related to an increase of the magnetic moment, 
with the intensity being proportional to the square of the moment. 
Therefore, the observed intensity increase of a factor of 4.6 for {\Bc} at 6.6~T implies that the AFM moment has more than doubled compared to that at zero field, while the moment only increases by 10\% for {\Ba}. 
Namely, the AFM moment increases monotonously from 0.17 to 0.36~\mub\ as the field \Bc\ goes from 0 to 6.6~T. 
If, in addition, a moment reorientation away from the basal plane with field would be involved in the observed field response, the increase in the moment size would be even larger than our estimate.
At the inflection points in the isotherms, an increased rate of the AFM moment with field changes its slope.
We cannot at this point conclude whether the boundary is a real transition or a cross-over, and further investigations are necessary to reveal the nature of this anomaly.

In order to get insights into the microscopic mechanism of the moment enhancement, the magnetic field dependence of the inelastic response near the AFM zone centre for \Bc\ was investigated.
Figure~\ref{f5} displays inelastic neutron scattering spectra at (-0.05~0~$\frac{1}{2}$) measured for fields of 0 and 3.8~T at $T=50$~mK.
In contrast to the strong enhancement of the AFM Bragg peak intensity by a factor of 2.5 at 3.8~T, 
the inelastic spectrum at this field is basically unchanged to that in zero field.
A damped spin-wave excitation at around 0.6~meV at 0~T remains at the same energy with slightly lower intensity at 3.8~T.
The low-energy response below this excitation is unchanged as well.

The effect of magnetic field on the AFM state has not been studied extensively. 
1/$T_1T$ measured at the Pt site shows notable magnetic field dependence, which suggests that the AFM ground state is sensitive to the field.\cite{Yogi2004}
A slight increase of {\tn} with magnetic field particularly for {\Bc} indicates that the AFM order is stabilized with fields,\cite{Takeuchi2007} which is consistent with the observed field-induced increase of the AFM moment.
Although the AFM moment is strongly enhanced by magnetic field, the ordered magnetic moment of 0.36~{\mub} at 6.6~T is still substantially reduced from that expected from the reported ground state doublet.
Since the ground state is well separated from the first excited doublet at 15~meV,\cite{CePt3Si_TWillers_01}
the observed field evolution of the AFM moment is dominated by the ground-state doublet,
and the possibility of a higher order multipole or a mixing with a crystal field excited state can be excluded.\cite{CePt3Si_TWillers_01} 
This is experimentally confirmed by the fact that no apparent change was detected in the magnetic excitations. 
On the other hand, the Kondo interaction is a likely mechanism for reducing the ordered moment in zero field.
Therefore, a reduction of the Kondo effect by the applied field could lead to an increase of the magnetic moment.
However, we have not found any evidence for a suppression or strong reduction of the Kondo interaction.
Furthermore, in heavy-fermion antiferromagnets with similar $T_{\rm K}$, such as CeIn$_3$\cite{Knafo2003} and CePd$_2$Si$_2$,\cite{VanDijk2000}
a magnetic field does not enhance but rather suppress AFM moments, as in most antiferromagnets.

The application of a magnetic field results only in exceptional cases in the  stabilization of AFM order. 
One example can be found in the case where AFM order competes with a SC state.
The magnetic field is expected to suppress the SC state, which concomitantly leads to a recovery of AFM order.
This mechanism is evoked to describe the field-induced enhancement of the AFM intensity in the prototypical heavy-fermion superconductor CeCu$_2$Si$_2$.\cite{Faulhaber2007,Stockert2010}
However, this scenario is not applicable to {\cp3s} where the coupling between the AFM and SC states is weak.
Indeed, the AFM peak intensity in {\cp3s} is suppressed only a few percent at most on passing through {\tsc}, while the long range AFM order in CeCu$_2$Si$_2$ disappears below {\tsc}\cite{Stockert2006}.  
A direct evidence is that the field-induced intensity increase of the magnetic Bragg peaks is not affected by the superconducting critical field {\bc2} in \cp3s\ and is found even well above {\tsc}, as shown in Fig.~\ref{f3}.
This observation also excludes the possibility of a magnetic contribution induced in a vortex lattice.\cite{Lake2001}
Hence, the field-induced enhancement in {\cp3s} is not associated with superconductivity.
Furthermore, the field-independence  of the inelastic spectra also excludes as possible origin the suppression of spin fluctuations near the AFM zone center.
In fact,  inelastic neutron scattering measurements near {\qaf} in zero field have shown that the low energy quasielastic fluctuations observed above {\tn} disappears at {\tn}.\cite{Fak2008}
Therefore, the field-induced enhancement does not arise from the magnetic response at {\qaf}, showing a distinct difference to other unconventional superconductors, such as CeCu$_2$Si$_2$.

Another possibility is that there exists a missing magnetic response below {\tn} in {\cp3s}.
The field-induced enhancement of the moment amplitude would then be associated with a suppression of this ``hidden'' magnetic spectral weight, 
which could also be associated with a suppression of quasielastic fluctuations below {\tn} in zero field, a behavior quite different from other heavy-fermion antifferromagnets.
In fact, NMR suggests a multiband nature in {\cp3s}.\cite{Mukuda2009a}
In addition, theoretical studies suggest the presence of several maxima in the susceptibility in the ($h~k$~0) plane  that reflects the presence of antisymmetric spin-orbit coupling originating from the lack of inversion symmetry,\cite{Takimoto2008,Yanase2008,Takimoto2009}  which leads to an anomalous spin susceptibility.\cite{Fak2014}
However, no experimental evidence for an additional inelastic magnetic response has been found, despite extensive surveys of large regions in $Q$-${\omega}$ space.
Further work to map the inelastic magnetic response in other regions of reciprocal space is highly desired to search for the hypothetical ``hidden'' magnetic spectral weight and to deepen our understanding of {\cp3s}. 

In conclusion, using elastic neutron scattering, we have observed an unusually strong field-induced enhancement of the AFM Bragg peak intensity for a field \Bc, which we interpret as an increase of the ordered AFM moment by the external field rather than a moment reorientation.
This leads to more than a doubling of the AFM moment at 6.6~T, while for \Ba\ we only observe a minor increase of the moment size, of the order of 10\%.
Our results reveal that the character of the AFM state in {\cp3s}, which coexists with superconductivity,  is markedly different from other unconventional superconductors

The authors would like to thank  S. Wakimoto, M. Yogi, Y. Tokunaga and T. Takimoto for stimulating discussions.
F. Honda, Y. Shimojo, P. Steffens and the sample environment teams at BER-II and ILL are acknowledged for technical assistance on neutron scattering experiments. 
This work was supported by a Grant-in-Aid for Scientific Research on Innovative Areas "Heavy Electrons" (No. 21102524), and on Priority
Areas of "New Materials Science Using Regulated Nano Spaces" (No. 22013022), of The Ministry of Education, Culture, Sports, Science, 
and Technology, Japan, and for Young Scientist (B) (No. 23740247) and Scientific Research (C) (No. 2454036) from the Japan Society for the Promotion of Science.

\bibliography{CePt3Si3_kanekok}
\end{document}